\documentclass[preprint,12pt]{elsarticle}
\usepackage{amssymb}

\newcounter{bla}

\journal{Computer Physics Communications}

\usepackage{algorithm,algorithmic}
\usepackage{subfig}
\usepackage[switch, displaymath, mathlines]{lineno}
\usepackage{siunitx}
\usepackage{dcolumn}
\usepackage[version=4]{mhchem}
\usepackage{url}

\newcolumntype{d}[1]{D{.}{.}{#1}}
\newcommand{\tabcenter}[1]{\multicolumn{1}{l}{#1}}

\begin{document}
\begin{frontmatter}
%
\title{ 86 PFLOPS Deep Potential Molecular Dynamics simulation of 100 million atoms with \textit{{ab initio}} accuracy}

\author[a]{Denghui Lu}
\author[b]{Han Wang}
\author[a]{Mohan Chen}
\author[c]{Jiduan Liu}
\author[d,e]{Lin Lin}
\author[f]{Roberto Car}
\author[f]{Weinan E}
\author[d]{Weile Jia \corref{author}}
\author[f]{Linfeng Zhang \corref{author} }

\address[a]{CAPT, HEDPS, College of Engineering, Peking University}
\address[b]{Laboratory of Computational Physics, Institute of Applied Physics and Computational Mathematics}
\address[c]{Peking University}
\address[d]{University of California, Berkeley }
\address[e]{Lawrence Berkeley National Laboratory}
\address[f]{Princeton University}

\cortext[author] {Corresponding author.\\\textit{E-mail address:} jiaweile@berkeley.edu,linfengz@princeton.edu}

\begin{abstract}
We present the GPU version of DeePMD-kit, which, upon training a deep neural network model using {\it ab initio} data, can drive extremely large-scale molecular dynamics (MD) simulation with {\it ab initio} accuracy. 
Our tests show that for a water system of $12,582,912$ atoms, the GPU version can be $7$ times faster than the CPU version under the same power consumption. 
The code can scale up to the entire Summit supercomputer.  
For  a copper system of $113,246,208$ atoms, the code can perform one nanosecond MD simulation per day, reaching a peak performance of $86$ PFLOPS ($43\%$ of the peak). 
Such unprecedented ability to perform MD simulation with {\it ab initio} accuracy
opens up the possibility of studying many important issues in materials and molecules, such as heterogeneous catalysis, electrochemical cells, irradiation damage, crack propagation, and biochemical reactions. 
\end{abstract}

\end{frontmatter}

\newpage

{\bf NEW VERSION PROGRAM SUMMARY}\\
\\
\begin{small}
\noindent 
{\em Program Title: DeePMD-kit}                                          \\
{\em CPC Library link to program files:} (to be added by Technical Editor) \\
{\em Developer's repository link:}  https://doi.org/10.5281/zenodo.3961106\\
{\em Code Ocean capsule:} (to be added by Technical Editor)\\
{\em Licensing provisions:} LGPL  \\
{\em Programming language: C++/Python/CUDA}                                   \\
{\em Supplementary material:}                                 \\
{\em Journal reference of previous version:}*
  http://dx.doi.org/10.17632/hvfh9yvncf.1 \\
{\em Does the new version supersede the previous version?:}* 
  Yes.\\
{\em Reasons for the new version:*}
   Parallelize and optimize the DeePMD-kit for modern high performance computers. \\
{\em Summary of revisions:}*
  The optimized DeePMD-kit is capable of computing 100 million atoms molecular dynamics with \textit{ab initio} accuracy, achieving 86 PFLOPS in double precision. \\ 
{\em Nature of problem(approx. 50-250 words):}
  Modeling the many-body atomic interactions by deep neural network models. Running molecular dynamics simulations with the models.\\
{\em Solution method(approx. 50-250 words):}
  The Deep Potential for Molecular Dynamics (DeePMD) method is implemented based on
the deep learning framework TensorFlow. Standard and customized TensorFlow operators are optimized for GPU. Massively parallel molecular dynamics simulations with DeePMD models on high performance computers are supported in the new version.\\

* Items marked with an asterisk are only required for new versions of programs previously published in the CPC Program Library.\\
\end{small}

\newpage

%
\newcommand{\Or}{\mathcal{O}}
\newcommand{\jump}[1]{\big[\hspace{-0.7mm} \big[ #1 \big]
  \hspace{-0.7mm} \big]}
\newcommand{\mean}[1] {\big\{ \hspace{-0.7mm} \big\{ #1 \big\}
  \hspace{-0.7mm} \big\}}
\newcommand{\abs}[1]{\left\lvert#1\right\rvert}
\newcommand{\norm}[1]{\left\lVert#1\right\rVert}
\newcommand{\average}[1]{\left\langle#1\right\rangle}
\newcommand{\bra}[1]{\langle#1\rvert}
\newcommand{\ket}[1]{\lvert#1\rangle}
\newcommand{\mc}[1]{\mathcal{#1}}
\newcommand{\DG}{\mathrm{DG}}
\newcommand{\ud}{\,\mathrm{d}}

\newcommand{\bd}[1]{\boldsymbol{#1}}
\newcommand{\wt}[1]{\widetilde{#1}}
\newcommand{\wh}[1]{\widehat{#1}}
\newcommand{\wb}[1]{\overline{#1}}

\newcommand{\onlinecite}[1]{\citenum{#1}}
\newcommand{\bvec}[1]{\mathbf{#1}}
\newcommand{\vr}{\bvec{r}}
\newcommand{\vR}{\bvec{R}}
\newcommand{\I}{\mathrm{i}} 

\newcommand{\vace}{\widetilde{V}_{\mathrm{X}}}
\newcommand{\kace}{K^{\mathrm{ACE}}}
\newcommand{\ext}{\mathrm{ext}}
\newcommand{\Hxc}{\mathrm{Hxc}}
\newcommand{\X}{\mathrm{X}}
\newcommand{\xc}{\mathrm{xc}}
\newcommand{\eff}{\mathrm{eff}}

\newcommand{\REV}[1]{{#1}}

\renewcommand{\algorithmicrequire}{\textbf{Input:}} 
\renewcommand{\algorithmicensure}{\textbf{Output:}}

\section{Introduction}
In recent years, there has been a surge of interest in using {\it ab initio} simulation tools for a microscopic understanding of various macroscopic phenomena in many different disciplines, such as chemistry, biology, and materials science. 
One of the most powerful tools has been the {\it ab initio} molecular dynamics (AIMD) scheme~\cite{car1985unified}:
By generating on-the-fly the potential energy surface (PES) and the interatomic forces from first-principles density functional theory (DFT)~\cite{hohenberg64,kohn1965self} during molecular dynamics (MD) simulations,  
it is possible to obtain an accurate description of the dynamic behavior of the system under study at the atomic level.
However, due to the complexity associated with DFT, the spatial and temporal scales accessible by AIMD have been limited. 
Most routine AIMD calculations can only deal with systems with hundreds of atoms on the time scale of picoseconds.  
Although many linear-scaling DFT methods have been developed~\cite{Goedecker1999,BowlerMiyazaki2012} and some of them have been implemented on high performance computing (HPC) architectures for large-scale atomic simulation with tens of thousands of atoms~\cite{wang2008linearly,eisenbach2009scalable}, they are mostly limited to insulating systems with relatively large band gaps.

For many problems of practical interests, such as  heterogeneous catalysis, electrochemical cells, irradiation damage, crack propagation in brittle materials, and biochemical reactions, etc., a system size of thousands to millions of atoms, or even larger, is often required. 
In these cases, one usually has to resort to empirical force fields (EFFs), currently the main driving force of large-scale MD.
In the past two decades, tremendous efforts have been made to develop parallel algorithms and softwares for EFF-based MD (EFFMD) on general purpose HPC machines~\cite{berendsen1995gromacs,hess2008gromacs,kale1999namd2,phillips2005scalable,case2005amber,plimpton1995fast,lagardere2018tinker,fitch2006blue,bowers2006scalable,glosli2007extending,germann2008trillion,hohnerbach2016vectorization,liu2018shift,sedova2018high,tchipev2019twetris,zhang2019sw_gromacs,li2019openkmc}. 
Representative examples include the optimization of the long-range electrostatic interaction~\cite{richards2009beyond,salomon2013routine,andoh2013modylas,mcdoniel2017lammps} and adapted MD algorithms for accelerators like GPU~\cite{stone2007accelerating,anderson2008general,salomon2013routine,pall2013flexible} and FPGA~\cite{waidyasooriya2016architecture,yang2019fully}. 
Besides general-purpose HPCs, there have also been constant attempts to build special-purpose hardware to boost the performance of MD simulation~\cite{narumi20001,taiji2003protein,narumi200655,shaw2008anton,shaw2014anton}. 
These attempts have made it possible to perform EFFMD for systems up to a spatial scale of sub-millimeters (twenty trillion atoms)~\cite{tchipev2019twetris} or a temporal scale of up to milliseconds~\cite{shaw2014anton}.
Unfortunately, the practical significance of these efforts is hindered by the limitation of the accuracy and transferability of the EFFs.
For example, it has been hard, if not impossible, to develop accurate and general-purpose EFF models for multi-element alloys.

Recent development of machine learning (ML) methods has brought new hope to addressing this problem and there have been a flurry of activities on ML-based models of the PES~\cite{behler2007generalized,bartok2010gaussian,rupp2012fast,chmiela2017machine,schutt2017schnet,smith2017ani,han2017deep,zhang2018deep,zhang2018end}. 
Despite the growing importance of the ML-based MD (MLMD),  publicly available softwares are still rare in comparison to the EFFMD. 
The few existing ones are mainly designed for MLMD running on desktop GPU workstations or on CPU-only clusters~\cite{khorshidi2016amp,yao2018tensormol,wang2018kit,abbott2019pes,lee2019simple,singraber2019library,quip}. 
\REV{For example, Lee et.~al.~reported an implementation of the Behler-Parrinello neural network potential interfaced with LAMMPS package. In a \ce{SiO2} system with 13,500 atoms, it scaled up to 80 CPU cores on a cluster with Intel Xeon E5-2650v2 CPUs~\cite{lee2019simple}. Singraber et.~al.~implemented the Behler-Parrinello neural network potential as a library and interfaced it with LAMMPS for molecular dynamics simulation. By using a water system with 2,880 molecules (8,640 atoms) as the test case, it was demonstrated that the implementation scaled to 512 CPU cores on a cluster with Intel Xeon
Gold 6138 CPUs~\cite{singraber2019library}. }
To the best of our knowledge, no attempt has been made to  implement and optimize MLMD to fully utilize the computational resources of modern heterogeneous supercomputers like Summit.
As a consequence,  although in principle MLMD makes it possible to achieve AIMD accuracy with EFFMD efficiency, 
this has not been realized in practice.

Among the various ML models proposed in the past few years, the Deep Potential (DP) scheme~\cite{han2017deep,zhang2018deep,zhang2018end} stands out as an end-to-end way of constructing accurate and robust PES models for a wide variety of systems.
This was made possible due to the smooth symmetry-preserving embedding sub-net in DP (in addition to the fitting net),  as well as the adaptive data generating scheme (in the framework of concurrent learning~\cite{zhang2020dp}) Deep Potential Generator (DP-GEN) \cite{zhang2019active}.
DP-based molecular dynamics (DeePMD) can reach the accuracy of AIMD while reducing its cost by several orders of magnitude.
Generalizations of the DP scheme have also made it possible to represent the free energy of coarse-grained particles~\cite{zhang2018deepcg} and various electronic properties~\cite{zhang2020dw,sommers2020raman,zepeda2019deep}.
In addition, an open-source implementation of DeePMD, named DeePMD-kit~\cite{wang2018kit}, has attracted researchers from various disciplines.
DP models have been used to study problems like first-order phase transitions~\cite{bonati2018silicon}, infrared spectroscopy and Raman spectroscopy~\cite{zhang2020dw,sommers2020raman}, nuclear quantum effects~\cite{ko2019isotope}, and various phenomena in chemistry~\cite{andrade2020free,zeng2019neural,chen2018deep} and materials sciences~\cite{dai2020theoretical,marcolongo2019simulating,wang2019deep,20JPCM-Liu}.

Nevertheless, the performances of DeePMD-kit and other DeePMD-based codes are limited by their sub-optimal implementation.
Although the training of DP models is rather efficient (typically less than one day on a single GPU card for most systems), extensive optimizations are required for model inference, namely to predict the energy and forces on-the-fly during an MD run, and to truly boost AIMD to large system size and long time scale.

To perform large-scale MD simulations, DeePMD-kit interfaces with LAMMPS~\cite{plimpton1995fast} and TensorFlow~\cite{abadi2016tensorflow}.
LAMMPS provides the basic infrastructure for MD, while TensorFlow provides a flexible toolbox for the deep learning part of DeePMD. 
In each MD step, DeePMD-kit retrieves atomic coordinates from LAMMPS that maintains the atomic information and the spatial partitioning of the system. 
Then environment matrices that describe the relative positions of atoms are computed from the coordinates.
In this step, the memory is accessed in a random order, which cannot be efficiently implemented by standard TensorFlow operators, so it is implemented by DeePMD-kit as a customized TensorFlow operator. 
Next, the environment matrices are converted to descriptors that describe the neighboring environment of atoms, and the descriptors are passed to a standard deep neural network (DNN) to produce atomic energies. 
This step is implemented by standard TensorFlow operators. 
Finally, the atomic energies and forces (obtained by back propagation) are returned to LAMMPS to update the atomic coordinates and momenta by numerical schemes.

The Summit supercomputer, which has a peak performance of $200$ PFLOPS (Peta floating point operations per second), provides us with an unprecedented opportunity to speedup DeePMD.
However, the original DeePMD-kit is not suitable for the heterogeneous architecture of Summit for the following reasons: (1) The environment matrix is only implemented on CPUs, this becomes the computational bottleneck when the descriptors and atomic energies are computed on GPUs. (2) Although  standard TensorFlow operators support GPU computation, the original DeePMD-kit can not assign multiple GPUs to multiple MPI processes in a massively parallel environment, thus only single GPU serial computation or multiple CPUs parallel computation are feasible. (3) The sizes of the DNNs in DP are relatively small, and the efficiency of the standard TensorFlow computational graph is relatively low.

To fully harness the power of Summit and future supercomputers, we need to address the following questions: (1) What is the best parallelization scheme for DeePMD-kit on a heterogeneous supercomputer like Summit? (2) How can we improve the efficiency of  DeePMD-kit on a GPU supercomputer for both customized and standard TensorFlow operators? (3) What is the scaling bottleneck of DeePMD-kit and how can we further improve its efficiency on architectures of future supercomputers? 
Furthermore, we would also like to understand: (1) What is the limit of  DeePMD-kit on Summit both in terms of system size and computational speed (time-to-solution)?  (2) What is the maximal achievable speedup factor of the GPU version of  DeePMD-kit versus the CPU version by using the same number of nodes or the same power consumption?  

The main contributions of this paper are: 
\begin{itemize}
    \item We find that DeePMD can use the same data distribution scheme of  EFFMD, and  parallelization is highly scalable on heterogeneous supercomputers. 
    \item By carefully optimizing the CUDA customized TensorFlow operators and re-constructing the architecture of the standard TensorFlow operators, DeePMD-kit can reach $43\%$ peak performance ($86$ PFLOPS) on Summit.
    \item By carefully analysing the scaling of DeePMD-kit, we identify the latency of both the GPU and network as the bottleneck of the current heterogeneous platform, which requires future improvements to push the limit of scales and applications that DeePMD-kit can handle.
    \item Weak scaling shows that the GPU version of DeePMD-kit can scale up to the entire Summit supercomputer, on a copper system 
    with $113$ million atoms. The strong scaling of a water system shows that DeePMD-kit can reach $110$ MD steps per second for a 4 million molecular water system with \textit{ab initio} accuracy.
    \item Our test results show that the GPU DeePMD-kit can be 39 times faster compared to the CPU version when using the same number of nodes, and 7 times faster under the same power consumption on Summit.
\end{itemize}

The rest of this paper is organized as follows: The Deep Potential algorithm is introduced in Section~\ref{sec:algorithm},
with implementation details  provided in Section ~\ref{sec:gpu}. The physical system and testing platform are presented in Sections~\ref{sec:system} and \ref{sec:machine}, respectively. Results  are discussed in Section~\ref{sec:result}, followed by a performance analysis in Section~\ref{sec:analysis}. Conclusions are drawn in Section~\ref{sec:conclusion}.  

\section{The Deep Potential model} \label{sec:algorithm}

The central quantity of an MD simulation is the PES $E$,
a function of the atomic coordinates $(r_1, \dots r_N) \in \mathbb R^{3N}$.
The DP model expresses $E$ as a sum of atomic contributions, i.e., ~$E = \sum_i E_i$.
The contribution $E_i$ from the atom $i$ depends only on $\mathcal R_i$, the local environment of $i$: $\mathcal R_i=\{r_{ij} : j \in L(i)\}$, where $r_{ij} = r_j - r_i$.
Here the neighbor index set $L(i)$ is defined by $\{j: \vert r_{ij} \vert \leq r_c\}$, and $r_c$ is a predefined cutoff radius.
In the DP model, $\mathcal R_i$ is first mapped via an embedding net onto a symmetry-preserving descriptor $\mathcal D$, and then $\mathcal D$ is mapped via a fitting network $\mathcal N$ to give $E_i$, i.e.,
\begin{align}\label{eq:ei}
  E_i = \mathcal N (\mathcal D (\mathcal R_i)).
\end{align}
Here the fitting net $\mathcal N$ is chosen to be a fully connected DNN with $l$ hidden layers:
\begin{align}\label{eq:fitting}
  \mathcal N(x) = \mathcal L^f_{l}\circ \cdots \circ \mathcal L^f_{1} (x),
\end{align}
where $\circ$ denotes the function composition.
Within each hidden layer, a skip connection between the input and the output is used,
\begin{align}\label{eq:fitting-detail}
   \mathcal L^f_{k} (x) = x + \tanh(x \cdot W^f_k + b^f_k),
\end{align}
with the weight $W^f_k$ being a square matrix and the bias $b^f_k$ being a vector with the same size as the input $x$.
The activation function $\tanh$ is applied component-wise.

The descriptor $\mathcal D$, which is required to preserve the translational, rotational and permutational symmetries, has the form
\begin{align}\label{eq:descrpt}
  \mathcal D(\mathcal R_i) = (\mathcal G_i^{<})^T \tilde{\mathcal R}_i (\tilde{\mathcal R}_i)^T \mathcal G_i,
\end{align}
where $\tilde {\mathcal R}_i\in\mathbb{R}^{N_m \times 4}$ is the environment matrix, and $N_m$ is the largest number of neighbors for all the atoms.
Each row of the environment matrix is a four dimensional vector: 
\begin{align}
  s(r_{ij})\times \Big(1, x_{ij}/\vert r_{ij}\vert, y_{ij}/\vert r_{ij}\vert, z_{ij}/\vert r_{ij}\vert \Big),
\end{align}
where $s(r_{ij}) = w(\vert r_{ij}\vert)/\vert r_{ij}\vert$ and $w(\vert r_{ij}\vert)$ is a gating function that decays smoothly from 1 to 0 at $\vert r_{ij}\vert=r_c$. 
The gating function ensures the smoothness of the environment matrix.
$(x_{ij}, y_{ij},z_{ij})$ are the Cartesian coordinates of $r_{ij}$.
If the number of neighbors of atom $i$ is less than $N_m$, the empty entries of $\tilde {\mathcal R}_i$ will be filled by zeros.
$\mathcal G_i\in\mathbb{R}^{N_m\times M}$ is called the embedding matrix, with each \REV{row} being an $M$ dimensional vector
\begin{align}
  \big( G_1(s(r_{ij})), \dots, G_{M}(s(r_{ij})) \big).
\end{align}
Here for each neighbor $j$, the input scalar $s(r_{ij})$ is mapped to the output $M$ dimensional vector $G = (G_1, \dots, G_M)$ via the so-called embedding net $G$, a DNN with the form
\begin{align}\label{eq:embedding}
  G(x) = \mathcal L^e_{m}\circ \cdots \circ \mathcal L^e_{1} \circ \mathcal L^e_{0}(x).
\end{align}
The first hidden layer is a standard feed forward network taking a scalar as input and outputting a vector of size $s_1$:
\begin{align}\label{eq:embedding-first}
  \mathcal L_0^e (x) = \tanh(x \cdot W_0^e + b_0^e),
\end{align}
where $W_0^e \in \mathbb R^{s_1}$ and $b_0^e \in \mathbb R$ denote the weight and bias, respectively. 
The rest of the hidden layers are expressed as
\begin{align}\label{eq:embedding-rest}
  \mathcal L_k^e (x) = (x, x) + \tanh(x\cdot W_k^e + b_k^e).
\end{align}
Here the output size is twice of the input size, i.e.,~$s_k = 2s_{k-1}$. The weight is a matrix of size $s_{k-1}\times s_k$ and the bias is a vector of size $s_k$.
$(x, x) \in \mathbb R^{s_k}$ denotes the concatenation of two $x\in \mathbb R^{s_{k-1}}$.
The only restriction imposed on the sizes of hidden layers is that the output size of the final layer should be identical to $M$, i.e.,~$s_m = M$.
In Eq.~\eqref{eq:descrpt}, the matrix $\mathcal G^<_i \in \mathbb R^{N_m\times M^<}$ with $M^{<} < M$  is a sub-matrix of $\mathcal G_i$ formed by taking the first $M^<$ columns of $\mathcal G_i$.

\emph{Remark 1.} The DP formulation \eqref{eq:ei} can be easily generalized to multi-component (with atoms of multiple chemical species) systems.
In this case, a fitting net $\mathcal N$ is built for each chemical species in the system,
i.e.,~$E_i = \mathcal N_{\alpha_i} (\mathcal D (\mathcal R_i))$,
where $\alpha_i$ denotes the chemical species of the atom indexed with $i$.
The chemical species of the neighbors of the atom $i$ are encoded in the descriptor~\eqref{eq:descrpt} by
separate embedding nets built for all possible combinations of the chemical species of two neighboring atoms,
i.e.,~$G^{\alpha_i,\alpha_j}(s(r_{ij}))$.
For example, for a system with 3 chemical species, 3 fitting nets and 9 embedding nets will be constructed.

\emph{Remark 2.} The force on atom $i$ is defined as the negative gradient of the total energy with respect to $r_i$:
\begin{align} \label{eq:dp-force}
  F_i = -\nabla_{r_i} E = -\sum_j \nabla_{r_i}  E_j ,
\end{align}
\REV{where the force components $\nabla_{r_i}E_j$ are calculated by the back propagation.
}

\emph{Remark 3.} For one evaluation of the DP model,
the fitting net is evaluated for each atom,
so the computational cost is of $\mathcal O(N)$.
The embedding net is evaluated for each pair of neighbors,
so the computational cost is of  $\mathcal O(N \times N_m)$.
The value of $N_m$ depends on the density of the system and $r_c$.
Usually $N_m$ is of the order $100 \sim 1000$.
Therefore, the evaluation of the embedding net is roughly two to three orders of magnitude more expensive than the fitting net. 


\section{Implementation}\label{sec:gpu}

\subsection{Parallelization}\label{sec:parallel}
\begin{figure}
  \centering
  \includegraphics[width=0.8\textwidth]{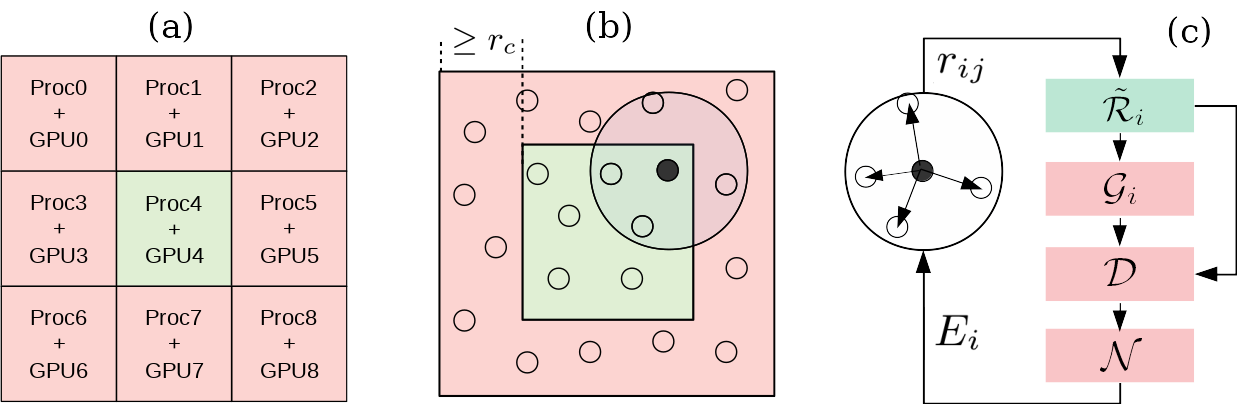}
  \caption{Data distribution and workflow of the DeePMD-kit.
    (a) Spatial subdivision of a system and the associated allocation of computational resources. Each sub-region is represented by a square.
    (b) The ghost region (red) for one sub-region (green). The open particles in the big circle with radius $r_c$ are the neighbors of the solid particle. The width of the ghost region should be equal to or larger than $r_c$. 
    (c) The single-atom DP workflow. The green step, i.e.,~the environment matrix $\tilde{\mathcal R}_i$, is implemented by a customized TensorFlow operator, while the red steps are implemented by standard TensorFlow operators (see the text for details). 
  }
  \label{fig:parallel}
\end{figure}

The DeePMD-kit takes advantage of the LAMMPS software package~\cite{plimpton1995fast} by replacing the \REV{short-range} EFF with the energy and forces derived from DP. Therefore, the data structure and parallelization strategy of LAMMPS are inherited by the DeePMD-kit code. 
\REV{More specifically, we provide a new \texttt{pair\_style} named \texttt{deepmd} by implementing DP as a new derived class of the base class \texttt{Pair}. It is worth noting that although DP is invoked by a "\texttt{pair\_style}", it is a multi-body interaction, which can be easily seen from the construction of DP in Sec.~\ref{sec:algorithm}.}

A two dimensional illustration of the data distribution in DeePMD-kit is shown in Fig.~\ref{fig:parallel}~(a). The physical system is divided into sub-regions, and then distributed among different computing units. 
For each sub-region, an extra ghost region of size larger than or equal to $r_c$ is needed to search neighbors of atoms, as shown in Fig.~\ref{fig:parallel}~(b). In each MD step, the single-atom DP workflow (Fig.~\ref{fig:parallel}~(c)) is conducted for each atom: First, the neighbor list is updated from the sub-region on the current computing unit, then the environment matrix $\tilde{\mathcal R}_i$ is computed from the neighbor list via a customized TensorFlow operator; Next, the atomic energy and force increments are evaluated through the DP model to update the atoms in the sub-region; Finally, the force increments in the ghost region are communicated among adjacent MPI processes, and global properties, such as energy, are communicated globally by {MPI\_Allreduce} operations. 
All of the positions and velocities of atoms are updated using the resulting forces according to a certain numerical scheme.

\REV{
\textit{Remark 4}.
The GPU version of LAMMPS takes advantage of the identity $\nabla_{r_i} E_j = - \nabla_{r_j} E_i$, which holds for most of the EFFs, so that the force of atom $i$ is given by $F_i = \sum_{j} \nabla_{r_j}  E_i $ according to Eq.~\eqref{eq:dp-force}. 
Therefore, all components of $F_i$ are computed on the computing unit that holds atom $i$, and the communications of force components are avoided. 
By contrast, the relation $\nabla_{r_i} E_j = - \nabla_{r_j} E_i$ dose not hold for DP as $i\neq j$, so we have to fallback to the CPU version of LAMMPS that is able to transfer back the component $\nabla_{r_i}  E_j$ computed on a non-native computing unit holding atom $j$ if it is in the ghost region.
}

\subsection{GPU Implementation and optimization} \label{sec:optimization}

\subsubsection{Naive GPU implementation}\label{sec:baseline}

The implementation of the DP model in DeePMD-kit is based on TensorFlow, a popular open-source software library for machine learning applications with GPU support~\cite{abadi2016tensorflow}. 
One common practice is to link the GPU supported TensorFlow to build the executable, so all DP operations implemented by standard TensorFlow operators (red boxes in Fig.~\ref{fig:parallel}(c)) are accelerated by GPU without additional effort. 
The testing results indicate that an overall 26.53 times of speedup can be achieved using a single NVIDIA V100 GPU compared to a single Intel Xeon Gold 6132 CPU core for a typical water system consisted of 12,288 atoms (4,096 molecules). 
We remark that throughout this section, the naive GPU implementation serves as the baseline of optimization.  The same water system is used for benchmarking purposes, and one MPI process using one thread is bound to a single GPU.


\subsubsection{Customized TensorFlow operators}\label{sec:customized_operators}

\begin{figure}[]
 \begin{center}
   {\includegraphics[width=0.8\textwidth]{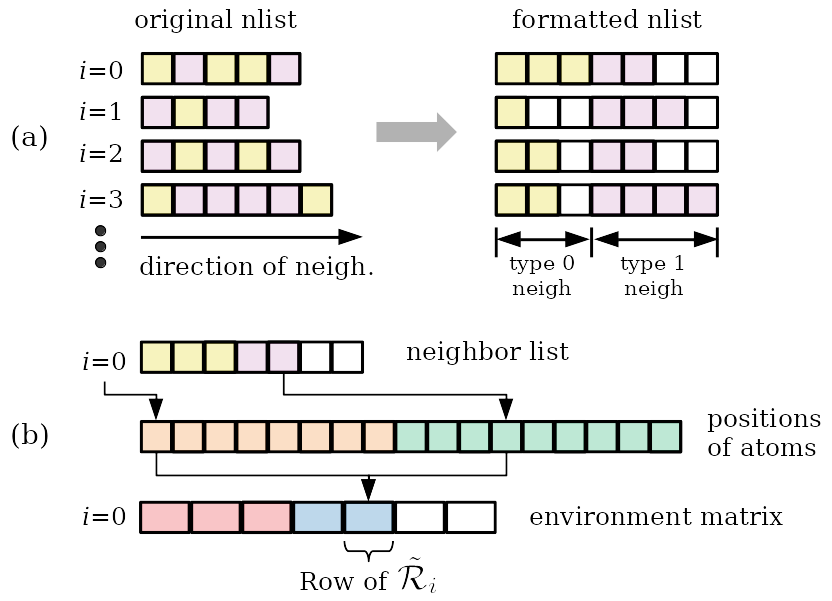}} 
   \end{center}
   \caption{
   Schematic plot of the computation of the environment matrix $\tilde{\mathcal R}_i$. (a) The first step: forming and formatting the neighbor list. The yellow squares stand for the neighbors of type 0, and the purple squares stand for the neighbors of type 1. The neighbors are first sorted according to their types (as shown in the figure), and then sorted according to their distance and their original atomic indices (not shown in the figure). The blank squares represent the padded positions in the neighbor list.
   (b) The second step: computing the environment matrix by using the formatted neighbor list, using the example of computing $\tilde{\mathcal R}_0$. 
   } 
   \label{fig:env_mat}
\end{figure}

\begin{algorithm}
    \caption{Formatting the neighbor list} 
    \label{alg:fmt_nlist}
    \begin{algorithmic}[1]
        \REQUIRE Atomic position $\{r_i\}$, the corresponding neighbor list $L(i,j)$ 
        \ENSURE  Formatted neighbor list $\tilde L(i,j)$
        \FOR{each $i \in [0,N_{l})$} 
            \FOR{each $k \in [0, L(i).\textrm{size})$} 
                \STATE $j = L(i, k)$
                \STATE $r_{ij} = r_j - r_i, 
                \quad \vert r_{ij}\vert = \sqrt{r_{ij}\cdot r_{ij}}$
                \STATE $S(i, k, 0) = \alpha(j),~ S(i, k, 1) = \vert r_{ij}\vert, ~S(i, k, 2) =j$
            \ENDFOR
            \STATE Sort the second dim with the third dim as key $S \rightarrow S^\ast$
            \STATE Pad the second dim $S^\ast \rightarrow S^{\ast\ast}$
            \STATE $\tilde L(i,:) = S^{\ast\ast}(i,:,2)$
        \ENDFOR
    \end{algorithmic}
\end{algorithm}

The customized TensorFlow operator for the environment matrix $\tilde{\mathcal R}_i$ is denoted by "Environment" and dominates the computational cost after linking the GPU TensorFlow library, because unlike the standard TensorFlow operators that support GPU, it only supports CPU in the original version of DeePMD-kit. 
The operator Environment includes two steps, formatting the neighbor list and computing the environment matrix by using the formatted neighbor list. 

The algorithm for formatting the neighbor list is shown in Alg.~\ref{alg:fmt_nlist}.  
The arbitrary ordered neighbor list of atom $i$ shown in Fig.~\ref{fig:env_mat}~(a) is sorted first based on the type of neighboring atoms, and then on the atomic distances $r_{ij}$.
In the case where two neighbors are of the same type and distance, the neighbor with a smaller atomic index is placed before the neighbor with a larger atomic index. 
The neighbors with different types in the neighbor list are then padded, so that they are aligned to the maximal number of neighbors of that type, as shown in Fig.~\ref{fig:env_mat}~(a). 
The reason for this operation is the following: in the computation of the embedding matrix, the neighbors of atoms $i$ are scanned over, and each row of the embedding matrix $\mathcal G_{i}$ is computed by passing $s(r_{ij})$ (the first element of the corresponding row of $\tilde{\mathcal R}_i$) to the embedding net $G^{\alpha_i,\alpha_j}$, which introduces a conditional branching according to the type of atom~$j$. 
Sorting and padding of the neighbor list avoids this unfavorable branching. 
In our GPU code, the construction of the neighbor list is still on CPU \REV{due to the problem of GPU LAMMPS for DP, as explained by Remark 4 of Sec.~\ref{sec:parallel}.}
In practice, the neighbor list is usually updated every 10 to 50 steps in an MD simulation, so our current implementation results in a satisfactory performance.


In order to efficiently format the neighbor list on the GPU, we perform the following optimization steps:


1. \textit {Naive CUDA customized kernels.}
The first step of optimization is to write a single CUDA customized kernel to accelerate the computation of Alg.~\ref{alg:fmt_nlist}. In this step, the first for loop (line 1) is unrolled with CUDA blocks and threads. Each CUDA thread is then responsible for calculating and sorting the neighbor list of a particular atom~$i$. 

2.\textit{ Converting array of structures (AoS) to structure of arrays (SoA).} 
A single element of the intermediate neighbor list $S$ is expressed by a structure (see Alg.~\ref{alg:fmt_nlist}). For example the $k$th neighbor of the $i$th atom $S(i, k)$ is a structure of three elements $(\alpha(j), \vert r_{ij}\vert, j)$, where $\alpha(j)$ and $j$ are integers and $\vert r_{ij}\vert$ is a floating point number.
Thus the corresponding GPU memory is not coalesced during the sorting procedure. One way of improving the GPU performance is to store the neighbor list as SoA instead of AoS. 
The SoA can improve the memory coalescing significantly, thus improving the performance of the CUDA kernel. 

3.\textit{ Unrolling of two for loops.} 
Two CUDA customized kernels are used to implement Alg.~\ref{alg:fmt_nlist} in this step. The first kernel is used to construct the intermediate neighbor list $S$ (line 1-6). In this implementation, the first and the second for loops are unrolled with CUDA blocks and threads respectively to further exploit the computing power of V100 GPU. Then the intermediate neighbor list is sorted and padded using a second kernel.

4. \textit{Compressing elements of the neighbor list to a 64 bit integer. }
The NVIDIA CUB library provides state-of-the-art and reusable software components for every layer of the CUDA programming model, including block-wide sorting.
To efficiently use the CUB library, we compress $S(i,k)$ 
into an \textsf{unsigned long long} number with the following equation: 
\begin{align}
  \tilde S(i, k) = \alpha(j) \times 10^{15} + \lfloor \vert r_{ij}\vert \times 10^{8} \rfloor \times 10^{5} + j
\end{align}
The 19 decimals of an \textsf{unsigned long long} integer is divided into 3 parts to store the neighbor list information: 4 decimal are used to store the atomic type of the neighbor atom ($\alpha(j)$), 10 decimals are used to store the distance of atom $i$ and its neighbor atom ($\vert r_{ij}\vert$), 5 decimals are used to store the atomic index of the neighbor atom ($j$). 
The range of all the three parts are carefully chosen to fulfill the restrictions that the total number of atom types is smaller than 1843, the cut-off radius is smaller than 100~\AA, and the number of neighbors is smaller than 100,000. 
These restrictions are rarely violated in typical MD simulations. 
The data compression is carried out before sorting, and a decompression procedure is needed afterwards. 
Both the compression and decompression are accelerated via CUDA customized kernels, and the corresponding computational time is negligible. 
We find that the compression reduces the total number of comparisons by half during the sorting procedure without deteriorating the accuracy of the result. 

\begin{figure}[]
  \begin{center}
  {\includegraphics[width=0.8\textwidth]{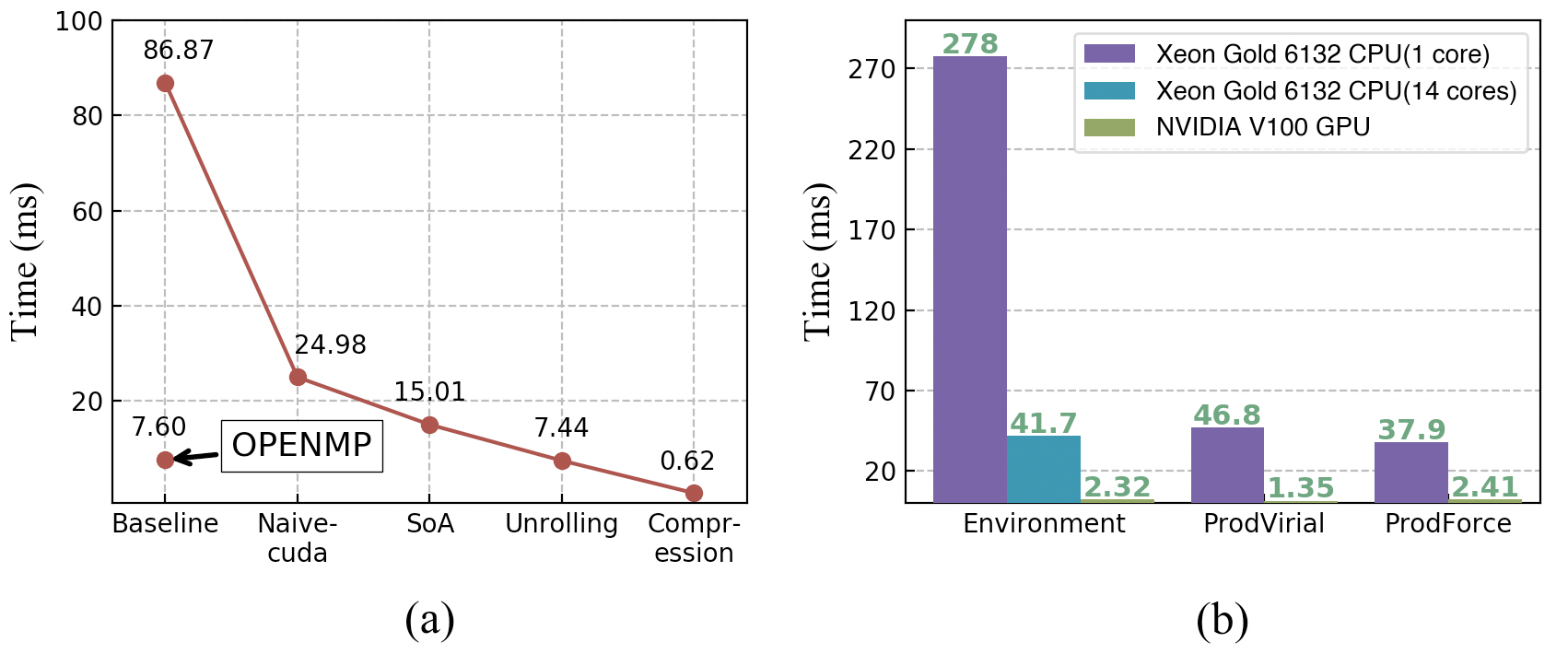}} 
  \end{center} 
  \caption{
    (a) Wall clock time versus different levels of GPU optimization for formatting the neighbor list, and (b) Performance of customized TensorFlow operators for a water system of 12,288 atoms.  
    \REV{The customized TensorFlow operators of the baseline code is implemented on the CPU. The wall clock time of the baseline is measured using both single CPU core and single Xeon Gold 6132 socket (14 cores with OpenMP threads set to 14). The GPU time is measured on single NVIDIA V100 GPU.}
  }
  \label{fig:environment-performance}
\end{figure}

Fig.~\ref{fig:environment-performance}~(a) shows the reduction of wall clock time associated with each stage of optimization. \REV{The baseline version is implemented on CPU, and tested with both single CPU core and single Xeon Gold 6132 socket with 14 cores by setting OpenMP threads to 14. }
We find that after all optimizations, the neighbor list formatting is accelerated by 141 times \REV{comparing to single CPU core and 12.25 times comparing to 14 CPU cores. We remark that when using both sockets with OpenMP threads set to 28, the performance is only slightly better than single socket due to the memory affinity}.

\begin{algorithm}
    \caption{Computing the environment matrix $\tilde{\mathcal R}$}
    \label{alg:environment}
    \begin{algorithmic}[1]
        \REQUIRE Atomic position $\{r_i\}$, formatted neighbor list $\tilde L(i)$
        \ENSURE Environment matrix ${\tilde{\mathcal R}_i}$
        \FOR{each $i \in [0,N_{l}]$} 
            \FOR{each $k \in [0, \tilde L(i).\textrm{size})$}
                \STATE $j = \tilde L(i, k)$
                \IF {$j$ is not a padded neighbor}
                    \STATE $r_{ij} = r_j - r_i, 
                    \quad \vert r_{ij}\vert = \sqrt{r_{ij}\cdot r_{ij}}$
                    \STATE $\tilde{\mathcal R}(i, k) = 
                    s(r_{ij}) \big(1, x_{ij}/\vert r_{ij}\vert, y_{ij}/\vert r_{ij}\vert, z_{ij}/\vert r_{ij}\vert \big)$
                \ELSE
                    \STATE $\tilde{\mathcal R}(i, k) = (0, 0, 0, 0)$
                \ENDIF
            \ENDFOR
        \ENDFOR    
    \end{algorithmic}
\end{algorithm}
The algorithm of the second step of the operator Environment, computing the environment matrix, is shown in Alg.~\ref{alg:environment} and graphically illustrated by Fig.~\ref{fig:env_mat}~(b). The formatted neighbor list is taken as input, and the corresponding  environment matrix is built based on line 6 in Alg.~\ref{alg:environment}. It is noted that the padded neighbors are skipped in the computation, and the corresponding places of the environment matrix are filled with zeros.

The optimization for the computation of the environment matrix follows the optimization steps 3 of formatting the neighbor list. The {\bf for} loops in Alg.~\ref{alg:environment} (line 1 and 2) are unrolled with CUDA blocks and threads. Each thread only works on a specific ${i,j,k}$ to fully exploit the computing power of V100 GPU. Two extra TensorFlow customized operators, \textrm{ProdVirial} and \textrm{ProdForce}, are also accelerated with the same fashion. These operators are used to calculate the force and virial outputs after the executions of embedding net and fitting net.

Fig.~\ref{fig:environment-performance}~(b) shows the wall clock time of the customized TensorFlow operators. The testing results show that our GPU implementation achieves 120, 35 and 16 times of speedup for the \textrm{Environment}, \textrm{ProdVirial}, \textrm{ProdForce} operators, respectively. \REV{We remark that operators ProdVirial and ProdForce are not fully optimized in the OpenMP implementation in DeePMD-kit because of the atomic addition, thus only sequential results of these two operators are shown in Fig.~\ref{fig:environment-performance}~(b).}
It is noted that the time for GPU memory allocations and the CPU-GPU memory copy operations are not included in the tests. For the water system consisting of 12,288 atoms, the total execution time of all three customized operators reduced from 363~ to $\sim$6~ms, achieving a speedup of 60 times.  Since the customized operators take $76\%$ 
of the total time, the GPU version of DeePMD-kit gains a speedup of 4.0 compared to the baseline implementation.   

\subsubsection{Optimization of the embedding net} \label{sec:embedding-net}

The environment matrix is used to compute the embedding matrix and assemble the descriptor, and finally the atomic energy contribution is computed by the fitting net, which takes the descriptor as input. 
All these steps are implemented by the standard TensorFlow execution graph. 
As discussed in Remark 2 in Section~\ref{sec:algorithm},
the computational cost of the fitting net is of order $O(N_m)$, while the cost of the embedding net is of order $O(N_m \times N_l)$, where $N_l$ being the number of atoms in the computing unit and $N_m$ being the maximal number of neighbors of an atom.
After optimizing the customized TensorFlow operators,  about 85\% of the total execution time is spent on the embedding net, while only 6\% of the execution time is used in the fitting net in our benchmark system. 
Therefore, in this section, we benchmark and optimize the performance of the embedding net. 
The embedding net (Eq.~\eqref{eq:embedding}) is composed of several hidden layers. Except for the very first layer~\eqref{eq:embedding-first}, the successive layers~\eqref{eq:embedding-rest} output a vector that is twice as large as the input vector. 
Most of the computational cost is spend on the successive layers (Eq.~\eqref{eq:embedding-rest}) rather than the first layer (Eq.~\eqref{eq:embedding-first}).
Therefore, we focus our attention on the successive layers.

\begin{figure}[]
    \begin{center}
    {\includegraphics[width=0.8\textwidth]{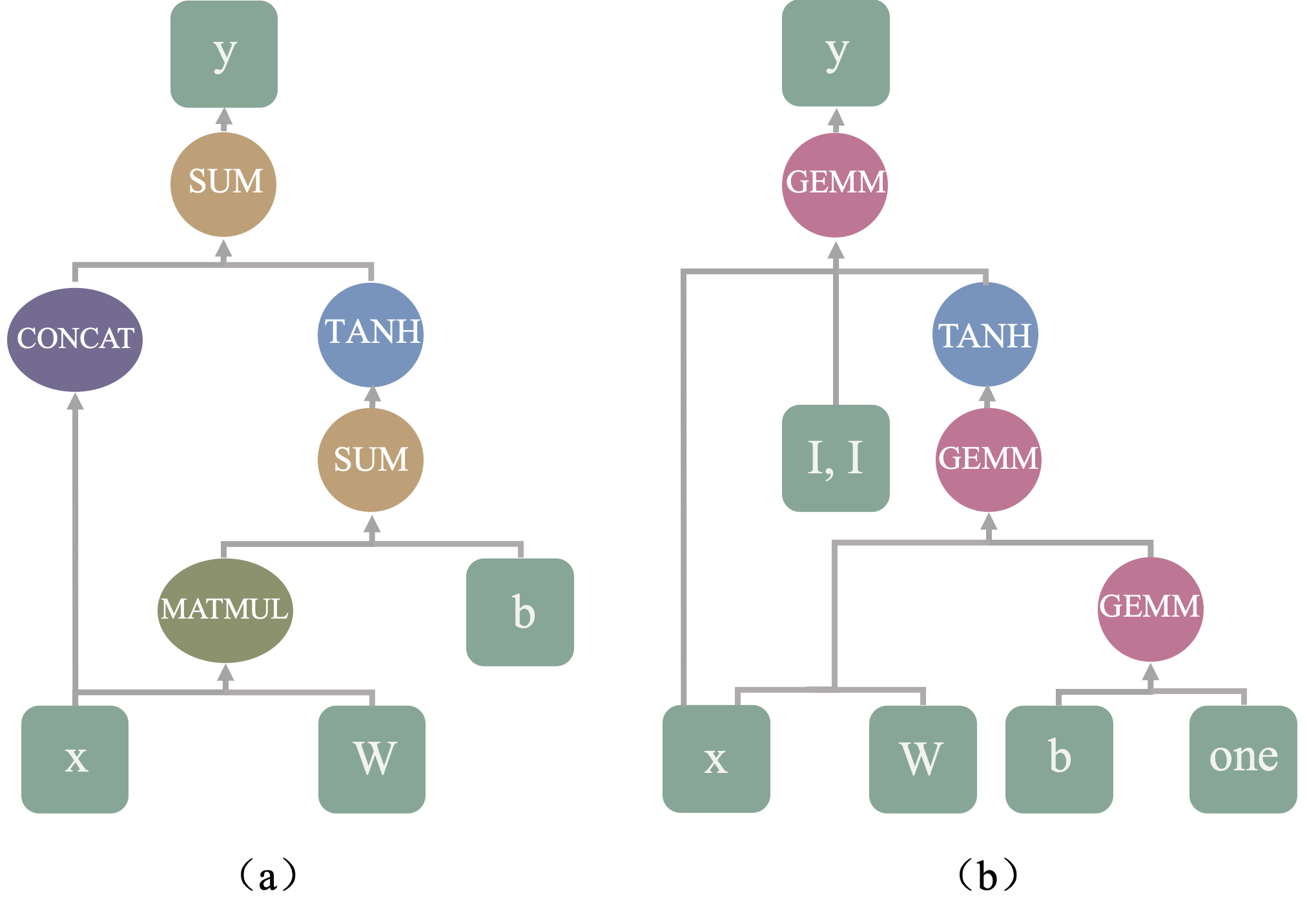}} 
    \end{center}
    \caption{
        Schematic plot of the execution graph of Eq.~\ref{eq:embedding-rest}. 
        (a) Implementation with the standard TensorFlow operators. 
        (b) Implementation with our optimized  TensorFlow operators. 
    }
    \label{fig:embedding-structure}
\end{figure}

The execution graph of Eq.~\eqref{eq:embedding-rest} with standard TensorFlow operators is presented in  Fig.~\ref{fig:embedding-structure}~(a). 
The TensorFlow operators such as the {MATMUL}, component-wise {SUM}, {TANH} and {CONCAT} are executed to perform the operations of matrix-matrix multiplication, summation, activation function, and concatenation, respectively. 
MATMUL and TANH are two of the most computationally intensive operators, 
and they can reach $72\%$ and $16\%$ of the peak on the GPU, respectively. 
Other operators such as CONCAT and SUM are bandwidth intensive with little floating point operations. Although linking to the GPU supported TensorFlow library provides considerable speedup compared to the CPU code, as shown in Section~\ref{sec:baseline}, our profiling results show that the total computational time is still dominated by those bandwidth intensive operators. 
For example, the computational time of CONCAT and SUM operators contributes 43\% of the total. 
Thus we identify these bandwidth-intensive operators as the ones that we make the greatest effort to optimize.

First, we notice that the summation and matrix-matrix multiplication are treated as two separated operators for evaluating $x \cdot W +b$ in the TensorFlow execution graph, as shown in Fig.~\ref{fig:embedding-structure}~(a). 
The MATMUL operator is invoked to calculate $x \cdot W$, where $x$ is a matrix of size  $376,832 \times 50$ (oxygen-hydrogen embedding) and $w$ is the weight matrix of size $50 \times 100$ in the benchmark system.
Next,  the SUM operator is called to add the bias $b$ to the resulting matrix $x \cdot W$. 
As shown in Fig.~\ref{fig:embedding-structure}~(b),
the MATMUL and SUM operators can be replaced by a single  CUBLAS GEMM call ($C = \alpha A \times B + \beta C$), which has both matrix-matrix multiplication and summation, thereby avoiding the corresponding SUM operator in the optimized implementation. 
It is noted that $b$ is a vector, and it is converted to a matrix format by multiplying with a transpose of the vector $one$. 
The wall clock for performing the SUM and MATMUL operators is reduced by $28\%$ after merging them into a single CUBLAS GEMM call.

Next, we move on to the optimization of  the CONCAT operator shown in Fig.~\ref{fig:embedding-structure}~(a). 
The CONCAT operator is performed to concatenate two $x$s to form $(x,x)$ in Eq.~\eqref{eq:embedding-rest}. 
The concatenation result, together with the result matrix of TANH operator, are summed up to produce the output of the embedding net.
In the standard TensorFlow execution graph, the CONCAT operator is implemented via the EIGEN library, which is a C++ template library for linear algebra.
In our optimized version, we replace the CONCAT operator with a matrix-matrix multiplication,
so that the following SUM operator (with the result of TANH) can be merged into one GEMM operator:
\begin{align}
\REV{
    (x, x) + \tanh(\cdots) 
    \rightarrow 
    \underbrace{ x\times (I, I) + \tanh(\cdots) }_{\textrm{GEMM}}.
}
\end{align}
It is noted that, in terms of performance, the matrix-matrix multiplication is marginally better than the implementation of CONCAT by EIGEN, and the main benefit comes from the removal of the SUM operator. 
The wall clock time of the CONCAT and SUM operators is reduced by $55\%$ after the optimization.

Last but not the least, we optimize the TANHGrad operator, which performs the derivation of $\tanh(x)$ in the backward propagation of the embedding net. It is noted that Fig.~\ref{fig:embedding-structure} only shows the forward propagation of the embedding net, and the TANHGrad operator is not included. 
However, in each MD step, both forward and backward propagation of the embedding net are executed. 
Noticing that the derivative of $\tanh(x)$ is also a function of $\tanh(x)$, i.e.,~$\frac{d}{dx}\tanh(x) = 1 - {\tanh^2(x)}$, 
we merge the TANH and TANHGrad operators by implementing both functions in the same CUDA customized kernel. 
Our testing results show that $37\%$ of the execution time is saved for the TANH and TANHGrad operators after optimization.

With all the optimizations above, an overall speedup factor of $1.18$ is achieved compared to the results in Section~\ref{sec:customized_operators}, and the cost of the matrix-matrix multiplication changes from $30\%$ to $61\%$ of the total execution time in the benchmark system.

\subsubsection{GPU memory accommodation} \label{sec:mem}
The memory footprint of the GPU version of DeePMD-kit sets the limit of the system size, since each NVIDIA V100 GPU on the Summit supercomputer only has 16 GB memory.
In the GPU code, the most memory demanding part is the embedding matrix $\mathcal G$. The number of floating point numbers to store one embedding matrix is approximately ${N_l \times N_m \times M}$.
Here $N_l$ is the number of atoms residing on the GPU, $N_m$ is the maximal number of neighbors, and $M$ is the width of the output layer of the embedding net. 
Therefore, the GPU memory requirement is not only restricted by the size of the network~($M$), but also related to the number of neighbors included in the neighbor list. 
In the execution, three layers of embedding net are used, and the output matrix size is twice of the input matrix. 
In the last layer, the sizes of both the output matrix and its derivative are ${N_l \times N_m \times M}$. 
An extra matrix of size ${N_l \times N_m \times M}$ is used to perform the concatenation operation. 
Therefore, a total of 4.5 copies of the embedding matrix are needed in the DeePMD-kit calculations as the equation shown below:
\begin{align}\label{eq:memory_usage}
     4.5\times N_l \times N_m \times M \times \texttt{sizeof}(data\_type)
\end{align}
For a typical system, such as the water and copper systems that will be discussed later, the memory requirement grows linearly with the number of atoms. 
Note that $N_m$ is usually of the order of a hundred, and $M$ is usually 100 in practice. 
For example, if we take $N_l = 25,000$, $N_m = 138$, $M = 100$ and \textit{data\_type = double}, the memory usage of $\mathcal{G}$ reaches 12.42~GB. 
This estimate can be verified by the numerical results in Section~\ref{sec:result}.

\section{The physical system} \label {sec:system}

\begin{figure}
    \begin{center}
    {\includegraphics[width=0.75\textwidth]{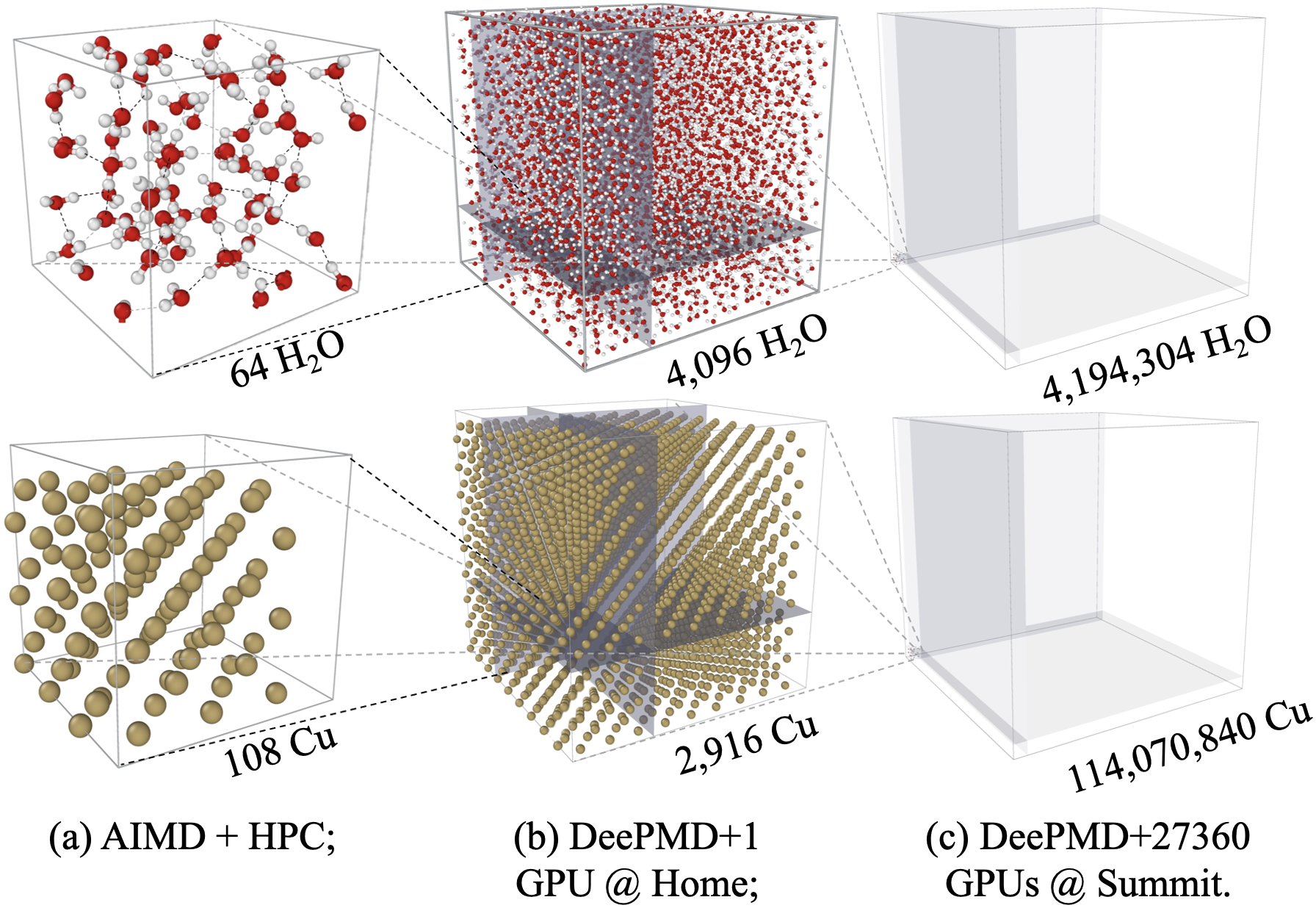}} 
    \end{center}
    \caption{Schematic illustration of two systems, water and copper, tested in this study. Shown in the figure are the system sizes accessible on different machines using different methods.
    (a) Sizes accessible by a typical AIMD simulation on HPC;
    (b) Sizes accessible by a typical DeePMD simulation with a single GPU;
    (c) Sizes accessible by the current study, the simulations using 27,360 GPUs on summit.}
    \label{fig:system-size}
\end{figure}

As shown in Fig.~\ref{fig:system-size}, we use two representative examples, water and copper, to investigate  the performance of the GPU DeePMD-kit  software package. 
Water, despite its simple molecular structure, has an unmatched complexity in the condensed (liquid) phase,  as a result of the delicate balance between weak non-covalent intermolecular interactions, e.g. the hydrogen bond network and van der Waals dispersion, thermal (entropic) effects, and nuclear quantum effects.
Copper represents an important and yet relatively simple metallic system, well suited as a  benchmark.
The training data of the water and copper systems are describe in Refs.~\cite{zhang2018deep,zhang2018end}, and \cite{zhang2020dp}, respectively. 
The DP models for both systems share almost the same architecture: sizes of the embedding and fitting nets are $25\times 50 \times 100$ and $240\times240\times240$, respectively. 
The cut-off radii of water and copper systems are 6~\AA\ and 8~\AA, respectively, and the maximal numbers of neighbors are 138 and 500, respectively.
Extensive benchmarks and theoretical studies have been conducted using DeePMD-kit, thus the accuracy of the model is reasonably assured. 
As a result, we can focus on the computational performance of the MD simulations.

The strong scaling of GPU DeePMD-kit is tested using the water system  composed of 12,582,912 atoms (4,194,304 water molecules), while the weak scaling is investigated using the copper system with 4,139 atoms per GPU card.
\REV{The configuration of the water system is made by replicating a well equilibrated liquid water system of 192 atoms for $64\times 32 \times 32$ times. The configurations of the copper system are generated as perfect face-centered-cubic (FCC) lattice with the lattice constant of 3.634~\AA. The FCC unit cell is replicated by $384\times 384 \times 192$ times to generate the largest copper system (113,246,208 atoms) tested in this work.}
The MD equations are numerically integrated by the Velocity-Verlet scheme for 500 steps (the energy and forces are evaluated for 501 times) at time-steps of 0.5~fs and 1.0~fs, respectively. 
The velocities of the atoms are randomly initialized subjected to the Boltzmann distribution at 330~K. 
The neighbor list with a 2~\AA\ buffer region is updated  every 50 time steps. 
The thermodynamic data including the kinetic energy, potential energy, temperature, pressure are collected and recorded in every 20 time steps.


\section{Machine configuration}  \label {sec:machine}

\begin{figure}[]
  \begin{center}{\includegraphics[width=0.5\textwidth]{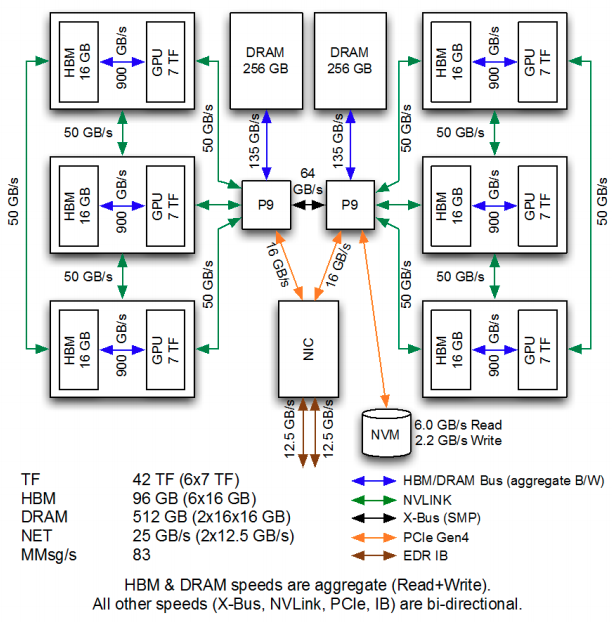}} \end{center}
  \caption{The architecture of a computational node on Summit. }
  \label{fig:summit}
\end{figure}

All numerical tests are performed on the Summit supercomputer. Fig.~\ref{fig:summit} shows the architecture of one of the 4608 Summit computing nodes. Each computing node consists of two identical groups, and each group has one IBM POWER 9 socket and 3 NVIDIA Volta V100 GPUs connected via NVLink with a bandwidth of 50 GB/s. Each POWER socket has 22 physical CPU cores and share 256 GB DDR4 CPU main memory, and each V100 GPU has its own 16 GB high bandwidth memory. The CPU bandwidth is 135 GB/s and GPU bandwidth is 900 GB/s. Each GPU has a theoretical peak performance of 7 TFLOPS double precision operations. The two groups of hardware are connected via X-Bus with a 64 GB/s bandwidth. The computing nodes are interconnected with a non-blocking fat-tree using a dual-rail Mellanox EDR InfiniBand interconnect with a total bandwidth of 25 GB/s.

In this paper, we utilize the MPI+CUDA programming model. In all the GPU tests, we use 6 MPI tasks per computing node (3 MPI tasks per socket to fully take advantage of both CPU-GPU affinity and network adapter), and each MPI task is bound to an individual GPU. 

\section{Numerical results} \label{sec:result}

\begin{figure}[]
  \begin{center}
  {\includegraphics[width=0.4\textwidth,angle=270]{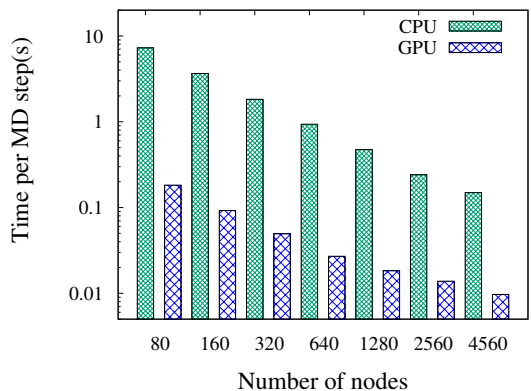}} 
  \end{center}
  \caption{Average wall clock time (log-scaled) of single MD step for a water system with 12,582,912 atoms using both CPU and GPU versions of the DeePMD-kit. 
  }
  \label{fig:cpu_gpu_compare}
\end{figure}

We compare the efficiency of the GPU version of DeePMD-kit to its CPU version for the water system with 12,582,912 atoms. In the CPU calculations, we utilize 42 MPIs per node to take full advantage of the Power 9 CPU sockets. In Fig.~\ref{fig:cpu_gpu_compare}, we \REV{measure} the wall clock time per MD step by \REV{averaging} over 500 MD steps using both the CPU and GPU versions of DeePMD-kit. All numerical experiments in this paper are performed using double precision due to the high accuracy nature of the DeePMD-kit code. 

\begin{figure}[]
  \begin{center}{\includegraphics[width=0.5\textwidth,angle=270]{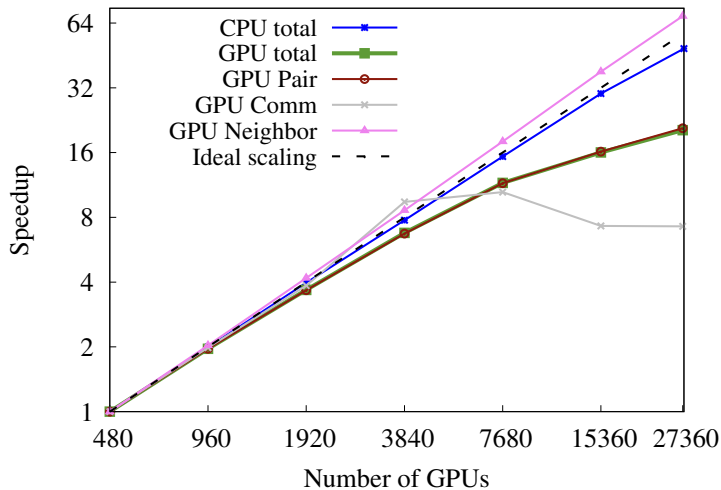}} \end{center}
  \caption{Strong scaling results of simulating a 12,582,912-atom water system with both CPU and GPU versions of the DeePMD-kit. The speedup is computed by setting the wall clock time for 80 nodes as the baseline. It is noted that the GPU baseline is 39 times faster compared to the CPU baseline as shown in Fig.~\ref{fig:cpu_gpu_compare}.
  } 
  \label{fig:strong_scaling}
\end{figure}

First, we compare the performance of the GPU version of DeePMD-kit to its CPU version with the same number of nodes.
Note that the CPU version can accommodate bigger physical systems because the size of the CPU  memory per node (512 GB) is $5$ times bigger than that of the GPU (96 GB) as shown in Fig.~\ref{fig:summit}. 
However, in terms of computational speed, testing results indicate that the GPU version can be 39 times faster on 80 Summit nodes (480 V100 NVIDIA GPUs against 3,360 POWER 9 CPU cores). 
The speedup factor decreases to $16$ when $4,560$ nodes are used (27,360 GPUs against 191,520 CPU cores). 
The decrease of the speedup factor is due to the fact that, as shown in Fig.~\ref{fig:strong_scaling}, the CPU code has a better strong scaling compared to the GPU code.
It is also worth noting that the GPU version is already much faster than the CPU version in the baseline implementation: as shown in Fig.~\ref{fig:cpu_gpu_compare}, the GPU baseline is 39 times faster than that of the CPU code when using 3,360 CPU cores, and even faster than that of the CPU code on 4,560 nodes. 
A detailed discussion of the scaling will be presented in Section~\ref{sec:analysis}.

Next, we compare the GPU version to the CPU version under the same power consumption, which is particularly important for the upcoming exascale computing era. 
The power consumption of a single POWER 9 socket is 190 watts, and 300 watts for a single NVIDIA V100 GPU. Hence, the power consumption of a single CPU node with 2 POWER 9 CPU sockets is 380 watts,
while the power consumption of each GPU node with 6 NIVIDA V100 GPUs and 2 POWER 9 CPU sockets is 2,180 watts. 
80 GPU nodes on Summit has a power consumption of 174,400 watts, and that is equivalent to the power consumption of 459 CPU nodes. 
In our tests, the GPU version of the DeePMD-kit can be $7$ times faster compared to the CPU version under the same power consumption. 

Fig.~\ref{fig:strong_scaling} demonstrates the strong scaling of a 12,582,912-atom water system with respect to the number of nodes. For this system, we find our GPU implementation can perfectly scale up to 640 nodes (3,840 GPUs) with 3,276 atoms per GPU, and continue to scale up to the entire Summit supercomputer~(4,560 nodes with 27,360 GPUs) with 455 atoms per GPU. We remark that the strong scaling defines the speed of the MD simulation, i.e.,~the GPU code can finish 110 MD steps per second for the water system of 12,582,912 atoms (4,194,304 molecules) when scaled to 4,560 Summit nodes. This delivers a capability of simulating the water system for 4.8~ns (with a time steps of 0.5~fs) in one day.

\begin{figure}[]
  \begin{center}{\includegraphics[width=0.6\textwidth]{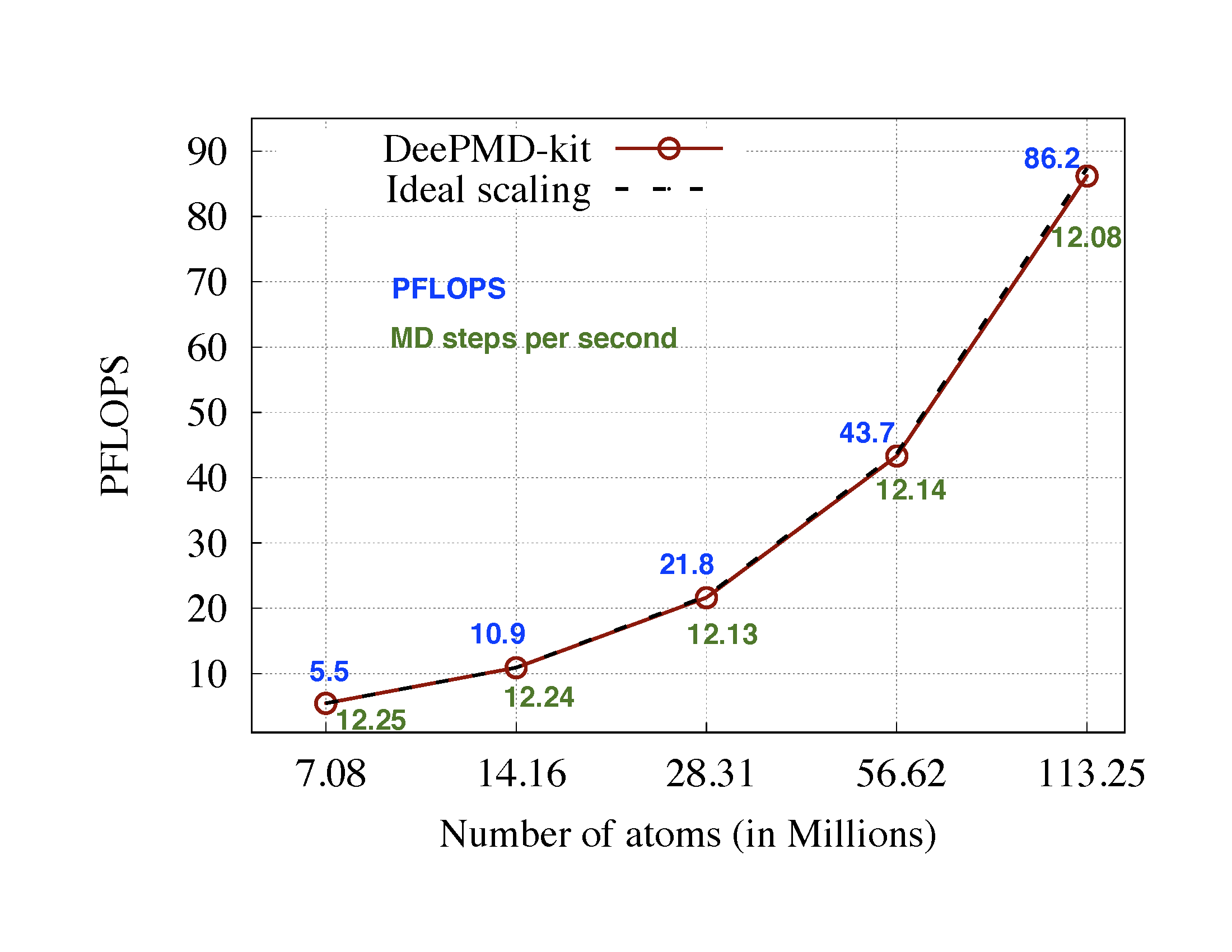}} \end{center}
  \caption{Weak scaling of the copper system. Each GPU holds 4,139 atoms, and the corresponding number of nodes scales from 285 to 4,560 throughout the tests.}
  \label{fig:weak_scaling}
\end{figure}

Fig.~\ref{fig:weak_scaling} shows the weak scaling of the GPU version of the DeePMD-kit for the copper systems. In the test, each MPI holds $4,139$ copper atoms on average. The number of GPUs scales from $1,710$ to $15,360$, and the corresponding number of atoms varies from $7,077,888$ to $113,246,208$, respectively. Our tests show that the GPU version of DeePMD-kit can achieve perfect scaling up to $4,560$ nodes. For the $113,246,208$ systems with copper atoms, we achieve $86.2$ PFLOPS with $4,560$ Summit nodes, reaching $43\%$ 
of the peak performance of Summit. Each MD step only takes $83$ milliseconds, therefore enabling one nanosecond simulation in one day.
A detailed discussion on the floating point operations per second (FLOPS) will be presented in the next section. 

\REV{
References~\cite{zhang2018deep,zhang2020dp} incorporate the "baseline implementation" of DeePMD-kit for MD simulations of water and copper systems, respectively, and have demonstrated that the accuracy of DeePMD is comparable to that of the AIMD simulation.
In this work, the optimized GPU DeePMD-kit uses the same floating point precision as the baseline implementation. The energy, force and virial predictions are found to be consistent with the baseline implementation up to 15, 10 and 13 digits, respectively. Therefore, the \textit{ab initio} accuracy of the GPU DeePMD-kit is warranted. 
}

\section{Performance Analysis} \label{sec:analysis}

\begin{table*}[ht]
\caption{
Wall clock time of the computationally intensive components for calculating 500 MD steps of 12,582,912 atoms of water system. All the testing results are in seconds. 
}
\centering
\begin{tabular}{l l l l l l l l }
\hline
\textbf{\#GPUs} & \tabcenter{480}  & \tabcenter{960}  & \tabcenter{1,920} & \tabcenter{3,840} & \tabcenter{7,680} & \tabcenter{15,360} &  \tabcenter{27,360} \\
\hline
Pair&88.4&45.03&24.08&13.13&7.66&5.46&4.25 \\
Comm&1.18&0.59&0.31&0.13&0.11&0.16&0.16 \\
Neighbour&2.39&1.18&0.57&0.28&0.13&0.06&0.03 \\
Others&0.32&0.30&0.12&0.09&0.08&0.08&0.09 \\
Total time &92.3&47.1&25.1&13.6&8.0&5.8&4.5 \\
\textbf{\#CPU cores}& \tabcenter{3,360} & \tabcenter{6,720} & \tabcenter{13,440} & \tabcenter{26,880} & \tabcenter{53,760} & \tabcenter{107,520} & \tabcenter{191,520~} \\
Total Time& 3632.8 & 1824.5 & 914.3 & 468.3 & 237.0& 120.8 & 74.5\\
\hline
\end{tabular}
  \label{table:total}
\end{table*}

\begin{figure}[]
  \begin{center}{\includegraphics[width=0.5\textwidth, angle=270]{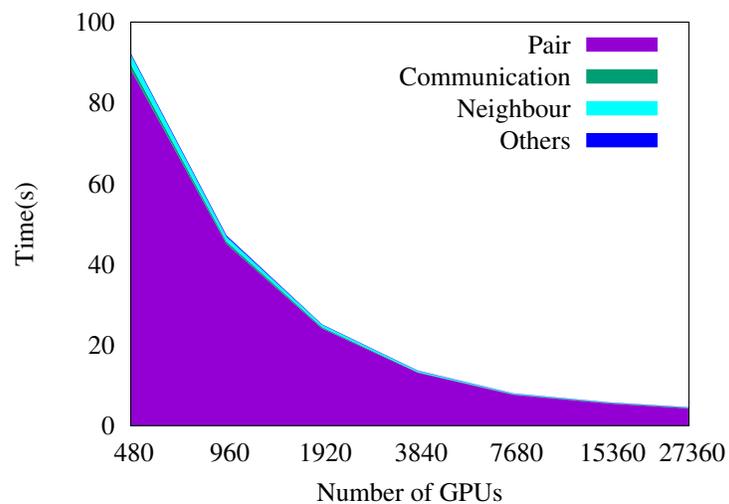}} \end{center}
  \caption{Wall clock time of 500 MD steps for a water system of 12,582,912 atoms.  }
  \label{fig:parts_scaling}
\end{figure}

In this section, we provide a detailed analysis for the GPU version of DeePMD-kit.
The total number of floating point operations (FLOP) for the $12,582,912$ atoms water system is $1.2483\times10^{17}$. This is collected from the CUDA profiling tool NVPROF. Although NVPROF only collects the FLOP number on the GPU, in our implementation, the CPU is only in charge of constructing and communicating the neighbor list and the corresponding FLOP number only accounts for less than $1\%$ of the total FLOP. The FLOPS is calculated by ${\text{(total FLOP)}} / { \text{(total time)}}$ and the corresponding efficiency is calculated via $\frac{ \textsf{FLOPS}} {\textsf{(number of nodes)} \times \textsf{43 TFLOPS}}$ (each V100 GPU has 7.0~TFLOPS, and each IBM Power 9 socket 515~GFLOPS, thus $7\times6+0.515\times2=43$~TFLOPS in total.). 
The efficiency of GPU version of DeePMD-kit is $38\%$ when using 480 GPUs, and decreases to $13\%$ when using $27,360$ GPUs for the water system with $12,582,912$ atoms, as shown in Fig.~\ref{fig:parts_scaling}. On the other hand, the weak scaling of the copper system shows the GPU version of DeePMD-kit achieves a peak performance of $86$ PFLOPS in double precision
with $4,560$ nodes on Summit ($43\%$ of the peak) when calculating $113,246,208$  copper atoms. 

\begin{figure}[]
    \begin{center}
    {\includegraphics[width=0.5\textwidth, angle=270]{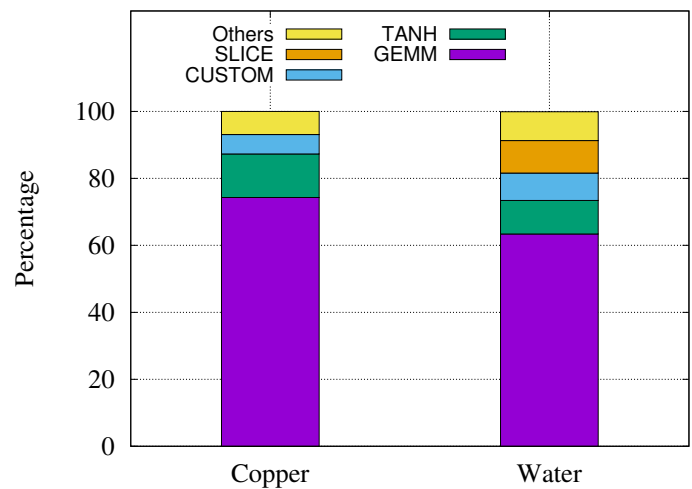}} 
    \end{center}
    \caption{
        Percentage of computational time by different TensorFlow operators for copper and water systems using the GPU version of DeePMD-kit.
        }
    \label{fig:proportions}
\end{figure}

We notice that the GPU version of DeePMD-kit shows better performance ($43\%$ of the peak) on the copper system than the water system ($36.8\%$ of the peak).  This is mainly due to two reasons: first, the average numbers of neighbors for each atom are $500$ and $138$ for the copper and water systems, respectively. Thus, the corresponding GEMM operation takes a larger proportion in the copper system compared to that of the water system. Secondly, since copper is a mono-species atomic system, no extra sorting and slicing in the computation of the embedding matrix is needed as discussed in Section~\ref{sec:gpu}. 
Fig.~\ref{fig:proportions} shows the proportion of different operations for both water and copper systems on the GPU. We find that the GEMM operator takes $92\%$ and $64\%$ of the GPU time for the copper and water system, respectively. 

The total computational time of the MD simulation can be divided into four parts: Pair, MPI Communication, Neighbor, and Others. The wall clock time for 500 steps of MD for each part is listed in Table~\ref{table:total} and shown in Fig.~\ref{fig:parts_scaling}.  
The corresponding strong scaling of the GPU DeePMD-kit is shown in Fig.~\ref{fig:strong_scaling}. 
The total wall clock time is dominated by the evaluation of the atomic energies and forces
, and denoted as ``Pair''. Therefore, the scaling of the Pair part is nearly the same as that of the total time in Fig.~\ref{fig:strong_scaling}.
The ``Comm'' part denotes the time used in updating the ghost region between adjacent MPI tasks. It scales with the number of GPUs when using less than 640 computing nodes, then becomes nearly a constant afterwards. This is because the communication time is gradually dominated by the network latency as the ghost region scales. 
The wall clock time for constructing the neighbor-list is labeled as "Neighbor". We notice that this operation shows superlinear scaling, which is attributed to both the reduction of the data size and better Cache hit ratio when more CPUs are used.
The "Others" part includes all other calculations such as the IO and the computations invoked by fixes, and it only contributes to less than $1\%$ of the total time, thus is negligible. 

We focus on analyzing the Pair part, which takes more than $93\%$ of the total time throughout the strong scaling tests for the $12,582,912$ atom water system. This part  includes the CPU-GPU memory copy operations and computation of the atomic energies and forces as discussed in Section~\ref{sec:optimization}. The efficiency of the Pair part is measured by the percentage of the peak performance as shown in Table~\ref{tab:data}. We find that the performance of this part highly depends on the data size. Note that when scaled up to 4,560 computing nodes, each GPU only holds 459 atoms on average, the total GPU memory usage is around 227 MB (from Eq.~\ref{eq:memory_usage}). The resulting data size cannot fully exploit the 7 TFLOPS computing power of the V100 GPU, which downgrades the efficiency. 
However, the GPUs are efficiently utilized when each GPU holds more than $3,200$ atoms.
The efficiency drops dramatically with less than 1,000 atoms per GPU. We also notice that the CPU-GPU memory copy slows down from $23.2$ GB/s with 480 GPUs to $4.7$ GB/s with 27,360 GPUs. Another reason for the drop of efficiency is the load imbalance when a large number of GPUs are used. For example, when scaled to 27,360 GPUs (4,560 nodes), the minimum number of atoms per GPU is 407 while the maximum is 505. The load imbalance leads to waits in the execution, and reduces the efficiency of the GPU. \REV{ The load balancing problem can be alleviated by dynamically redistributing the atoms onto the MPI tasks in the MD calculation \cite{bhatele2009dynamic,chen2010dynamic,brown2011implementing,glaser2015strong}.}

\begin{table}[ht]
\caption{\label{tab:data} Average number of atoms (per GPU), ghost atom number (per GPU) and FLOPS for a 12,582,912 atom water system.}
\begin{tabular}{ l r r r r r r r} 
 \hline
 \#GPUs & 480 & 960 & 1920 & 3840 &7680 & 15360 & 27360\\
 \hline 
 \#atoms  & 26214 & 13107 & 6553 & 3276 & 1638 & 819 & 459 \\
 \#ghosts  & 25566 & 16728 & 11548 & 7962 & 5467 & 3995 & 3039 \\
 \hline
 PFLOPS   & 1.35 & 2.65 & 4.98 & 9.16  & 15.63 & 21.66 & 27.51 \\
 \% \footnotesize{of Peak}   & 38.54 & 37.76 & 35.46 & 32.64 & 27.85 & 19.30 & 13.75 \\
\hline
\end{tabular}
\end{table}

The communication of the ghost region is performed with the adjacent MPI tasks, and the data size is listed in Table~\ref{tab:data}. The received size of a ghost region for each GPU from its neighboring MPI tasks is 25,566 (613 KB) when using 480 GPUs, and decreases to 3,039 (73 KB) when using 27,360 GPUs. Table~\ref{table:total} shows that the communication time decreases as the data size becomes smaller from 480 to 7,680 GPUs. Eventually, the communication time of the ghost region is dominated by the latency of the network, thus it stops scaling when using $15,360$ and $27,360$ GPUs in Table~\ref{tab:data}.

Collective MPI communication is also needed in obtaining the global properties for data IO during the simulation. Properties such as total energy, the stress, and the temperature, etc. are collected via \textrm{MPI\_Allreduce}. Since each of those properties is merely one double precision number, the \textrm{MPI\_Allreduce} operations are dominated by network latency. However, these latency can be a bottleneck in the extreme scale run if the physical properties are collected at every time step. By setting the output of the above mentioned properties to every 20 time steps, we find that the latency only accounts for less than $1\%$ of the total time. \REV{Since the latency is mainly caused by the implicit \textrm{MPI\_Barrier}, it can be further avoided by using the asynchronous \textrm{MPI\_Iallreduce} operation. }

\section{Conclusion} \label{sec:conclusion}

In this work, we propose the GPU adapted algorithms and re-implement the DeePMD-kit package on the heterogeneous supercomputer Summit. 


The weak scaling tests show that DeePMD-kit can scale up to $99\%$ of the Summit supercomputer, reaching a peak performance of $86.2$ PFLOPS ($43\%$ of the peak). For this particular system, each MD step  only takes 83 milliseconds, thereby enabling nanoseconds time scale simulation with \textit{ab initio} accuracy for the first time. For a typical water system consisting of $12,582,912$ atoms, our GPU code can scale up to 27,360 GPUs and run MD for 110 steps in one second. Compared to the CPU version, the GPU code is $16-39$ times faster when using the same number of nodes, and 7 times faster under the same power consumption. 

\REV{To study problems like heterogeneous catalysis, electrochemical cells, irradiation damage, crack propagation in brittle materials and biochemical reactions, the required system size for molecular simulation ranges from thousands to hundreds of millions of atoms.}
\REV{Traditionally these systems would be investigated with EFFMDs, from which scientific conclusions could not be solidly derived due to the limited accuracy of EFFs.
The unprecedented accuracy and efficiency of GPU DeePMD-kit, as realized in this work by integrating physical-based modeling, deep learning and optimized implementation on the world's largest supercomputer, open a new era of large scale molecular simulation, and will lead to groundbreaking scientific discoveries and innovations.
}

The success of our GPU code relies on: (1) adapting the data distribution of the classical MD software, (2) carefully optimizing the customized TensorFlow operators on GPU  (3) optimizing the standard TensorFlow operators on GPU. We remark that all these optimization techniques can be employed by other DP MD packages. We also analyze the scaling, and identify that the latency of both GPU and the network is the key for future improvement of exascale supercomputers to further accelerate the DP MD codes.  

Although we only demonstrate the optimization on the GPU Summit supercomputer, such strategies can also be applied to other heterogeneous architectures. For example, it can be easily converted to the Heterogeneous-compute Interface for Portability (HIP) programming model to run on the next exascale supercomputer Frontier, which will be based on AMD GPUs.

\section{Acknowledgement} \label{sec:ack}
Numerical tests were performed on the Summit supercomputer located in the Oak Ridge National Laboratory. 
The work of H. W. is supported by the National Science Foundation of China under Grant No. 11871110, the National Key Research and Development Program of China under Grants No. 2016YFB0201200 and No. 2016YFB0201203, and Beijing Academy of Artificial Intelligence (BAAI). This work was partially supported by the National Science Foundation
under Grant No. 1450372, No. DMS-1652330  (W. J. and L. L.), and by the Department of Energy under Grant No. DE-SC0017867 (L. L.). We thank a gift from iFlytek to Princeton University and the ONR grant N00014-13-1-0338 (L. Z. and W. E), and the Center Chemistry in Solution and at Interfaces (CSI) funded by the DOE Award DE-SC0019394 (L. Z., R. C. and W. E). The authors would like to thank Junqi Yin, Lin-Wang Wang for helpful discussions. 

%
\bibliographystyle{unsrt}
\bibliography{ref}{}
%

\end{document}